\documentclass{DISproc}

\begin{document}
\title{Hard Exclusive $\rho^0$-Meson Production at \mbox{COMPASS}}

\author{{\slshape Heiner Wollny for the COMPASS collaboration}\\[1ex]
CEA Saclay, 91191 Gif-sur-Yvette, France\\
}
\contribID{127}

\doi  

\maketitle

\begin{abstract}
New results for the transverse target spin azimuthal asymmetry
$A_{UT}^{\sin(\phi-\phi_S)}$ for hard exclusive $\rho^0$-meson production
on a transversely polarised $^6$LiD and NH$_3$ target will be presented. The
measurement was performed with the COMPASS detector using the
$160$\,GeV/$c$ muon beam of the SPS at CERN. The asymmetry is sensitive to the
nucleon helicity-flip generalised parton distribution $E$, which is
related to the orbital angular momentum of quarks in the nucleon.
\end{abstract}

\section{Introduction}
Hard exclusive vector meson production on nucleons is an important tool
to study strong interactions. Moreover, in Bjorken kinematics it
provides access to generalised parton distributions
(GPDs)~\cite{Mueller:1998fv,Radyushkin:1996ru,Ji:1996ek}. The GPDs contain a
wealth of information on the structure of the nucleon. In particular,
they embody both nucleon electromagnetic form factors and parton
distribution functions. Furthermore GPDs correlate longitudinal momenta
and transverse spatial position of partons referred to as 3-dimensional
nucleon tomography~\cite{Burkardt:2000za} .

At leading twist, vector meson production is described by the GPDs $H^f$
and $E^f$, where $f$ denotes a quark of flavour $f$ or a
gluon. The GPDs are functions of $t$, $x$ and $\xi$, where $t$ is the
squared four-momentum transfer to the nucleon, $x$ is the average and $\xi$
is half the difference of the longitudinal momenta carried by the struck
parton in the initial and final state. The GPDs $H^f$ describe the case
where the nucleon-helicity is conserved in the scattering process,
whereas the GPDs $E^f$ describe the case of nucleon-helicity flip. Hence, in the latter case angular momentum must be involved in order to conserve total angular momentum. The Ji relation, connects the total angular momentum of a given parton
species $f$ to the second moment of the sum of GPDs $H^f$ and
$E^f$~\cite{Ji:1996ek}:

\begin{equation}
J^f = \frac{1}{2} \lim_{t \to 0} \int^1_{-1} \text{d}x\, x \left[ H^f(x,\xi,t) + E^f(x,\xi,t) \right].
\end{equation}
This relation attracted much attention since it is the only known way to
constrain the quark angular momentum to the nucleon spin budget.  In
vector meson production on unpolarised nucleons the GPDs $E^f$ are
suppressed in COMPASS kinematics and one is only sensitive to the GPDs
$H^f$. However, in the cross-section for transversely polarised nucleons
the GPDs $E^f$ appear at leading twist in the azimuthal asymmetry
$A^{\sin(\phi-\phi_S)}_{UT}$. Here, the indices $U$ and $T$ refer to the
unpolarised beam and transversely polarised target, respectively. The
superscript $\sin(\phi-\phi_S)$ indicates the type of azimuthal
modulation of the cross-section, where $\phi$ is the azimuthal angle
between lepton scattering plane and the plane defined by the virtual
photon and the produced meson, and $\phi_S$ is the azimuthal angle of the
target spin vector relative to the lepton scattering plane. In this
paper the asymmetry $A^{\sin(\phi-\phi_S)}_{UT}$ for exclusive
$\rho^0$-meson production on transversely polarised deuterons and
protons is presented.


\section{Data sample and event selection}
The presented analysis is performed on data taken with the COMPASS
spectrometer~\cite{Abbon:2007pq} by scattering positive muons of
160\,GeV/$c$ from the CERN SPS off transversely polarised solid state
targets. The data taken in 2003 and 2004 with
a $^6$LiD target and the data taken in 2007 and 2010 with a NH$_3$
target were analysed. For the $^6$LiD material the average target polarisation is
about 0.5, while for NH$_3$ it is about 0.8. The target dilution factor
for exclusive $\rho^0$ production is typically 0.45 and 0.25 for the
$^6$LiD and NH$_3$ target, respectively. Before 2006 the target
consisted of two separated cells of equal length, which were oppositely
polarised. Since 2006 three target cells were used, where neighbouring
cells were oppositely polarised and the length of the two outer cells
matches the length of the middle one. This ensures a better balanced
acceptance for cells with opposite polarisation. Both target layouts
allow a simultaneous measurement of both target spin directions
compensating flux dependent systematic uncertainties. In addition, to
reduce systematic effects of the acceptance the polarisation of the
target cells were reversed about every week.

A new solenoid magnet installed during the shut down in
2005 increased the angular acceptance of the experiment from $\pm
70$\,mrad to the design value of $\pm 180$\,mrad.

Events in the DIS regime are selected by cuts on squared four-momentum
transfer $Q^2 > 1$\,(GeV$/c)^2$, on the fractional energy lost of the
muon $0.1 < y < 0.9$ and on the invariant mass of the $\gamma^*-N$
system $W > 5$\,GeV$/c^2$. An upper cut on $Q^2 < 10$\,(GeV$/c)^2$ is
applied to remove the region where the fraction of non-exclusive
background is large.  The interaction vertex is required to be inside
the polarised target material and the extrapolated track of the beam
muon must traverse the full length of the target, in order to ensure
equal flux in all target cells. The $\rho^0$-meson is detected via its decay into a
$\pi^+\pi^-$ pair, with invariant mass $0.5$\,GeV/$c^2 < M_{\pi\pi} <
1.1$\,GeV/$c^2$. Only events with an incident muon track, a scattered
muon track and two additional tracks with opposite charge are
considered. Suppression of semi-inclusive deep-inelastic (SIDIS)
production is achieved by a cut on the energy of the $\rho^0$-meson and on the
missing energy $E_{\text{miss}}=(M_X^2-M_P^2)/(2M_P)$, where $M_X$ is
the invariant mass of the undetected system and $M_P$ is the proton
mass:
$E_{\rho^0} > 15\,\text{GeV}$ and $-2.5\,\text{GeV} < E_{\text{miss}} < 2.5\,\text{GeV}$.
In order to suppress background of coherent production of
exclusive $\rho^0$ production on nuclei of the target a lower cut on the transverse momentum $p_T$ of the $\rho^0$-meson is applied:
$0.1\,(\text{GeV}/c)^2 < p_T^2 < 0.5\,(\text{GeV}/c)^2$ for $^6$LiD and $0.05\,(\text{GeV}/c)^2 < p_T^2 < 0.5\,(\text{GeV}/c)^2$ for NH$_3$, 
where the upper cut on $p_T^2$ is applied to remove the region, where non-exclusive production dominates.

\begin{figure}
\begin{center}
 \includegraphics[height=0.41\textwidth,trim= 0 0 0 0,clip,angle=-90]
        {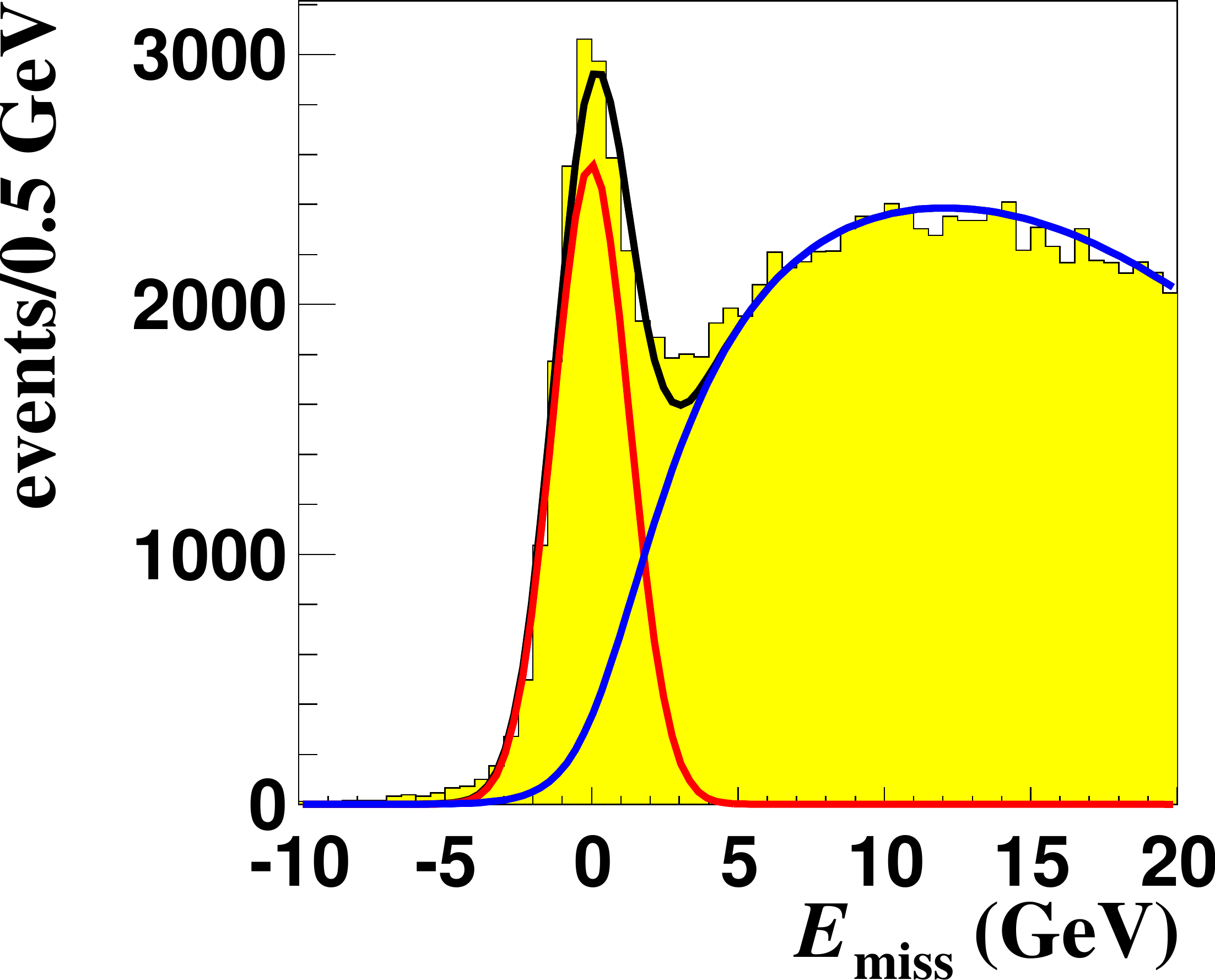}
\hspace{10mm}
 \includegraphics[height=0.41\textwidth,trim= 0 0 0 0,clip,angle=-90]
        {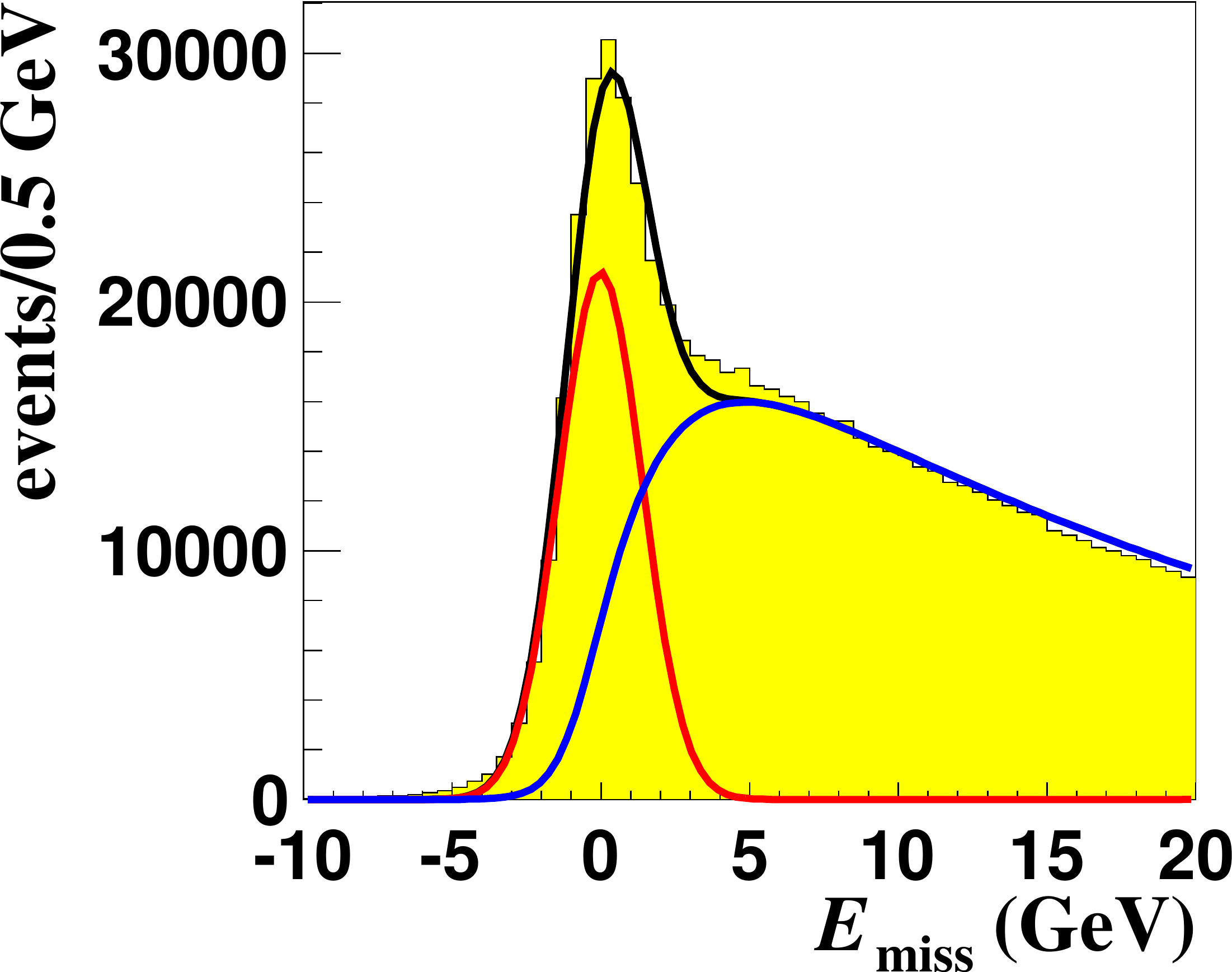}
\caption{Signal plus background fit to the $E_{\text{miss}}$ distribution for the $^6$LiD data (left) and NH$_3$ data (right) in the region $2.4$\,(GeV$/c)^2<Q^2<10$\,(GeV$/c)^2$ where the SIDIS background is the largest. The different shapes of the two distributions for large $E_{\text{miss}}$ is explained by the increased angular acceptance of the COMPASS detector due to the new target magnet installed in 2005.}
\label{PIC:EmissFit}
\end{center}
\end{figure}

After all cuts the final samples of incoherent exclusive $\rho^0$
production consist of about 97000 and 797000 events for the $^6$LiD and
NH$_3$ target, respectively. Still, about 20\,\% of the events originate
from SIDIS production. In order to correct for this contribution fits to
the $E_{\text{miss}}$ distribution are performed in each bin required
for the asymmetry extraction. This means in bins in $x_{Bj}$, $Q^2$, or
$p_T^2$ per target cell, and also in $\phi-\phi_S$ and according to the target spin orientation. For
the signal a Gaussian distribution is fitted. The shape for the
background is obtained from a parameterisation of SIDIS Monte Carlo (MC)
data generated with LEPTO binned in the same way as the real data,
except in polarisation state and $\phi-\phi_S$, since no polarisation
effects are simulated. The shape obtained from MC is fixed and only the
normalisation is fitted to the data. This is exemplarily shown in
Fig.~\ref{PIC:EmissFit} for both the $^6$LiD and NH$_3$ data in the
region $2.4$\,(GeV$/c)^2<Q^2<10$\,(GeV$/c)^2$ integrated over the target
cells, polarisation state and angle $\phi-\phi_S$. The resolution in
$E_{\text{miss}}$ is not sufficient to resolve $\rho^0$ production with
diffractive dissociation of the target nucleon. It is visible in
Fig.~\ref{PIC:EmissFit} right of the exclusive peak as a slight
enhancement over the SIDIS background. The amount was estimated using
Monte Carlo generated with HEPGEN~\cite{andrzej_hepgen} to be $\approx 14$\,\%. No attempt was made to remove this
background. This is motivated by HERA results for unpolarised protons,
which demonstrated that the angular distributions for
proton-dissociative production are consistent with those of exclusive
$\rho^0$ production~\cite{Breitweg:1999fm,Chekanov:2007zr,Aaron:2009xp}.
The asymmetry is extracted from the numbers of exclusive events, after 
subtraction of SIDIS background, using a binned maximum likelihood fit in 12 bins of $\phi-\phi_S$.

\section{Results}

\begin{figure}
\begin{center}
 \includegraphics[width=0.765\textwidth,trim= 0 35 0 0,clip]
        {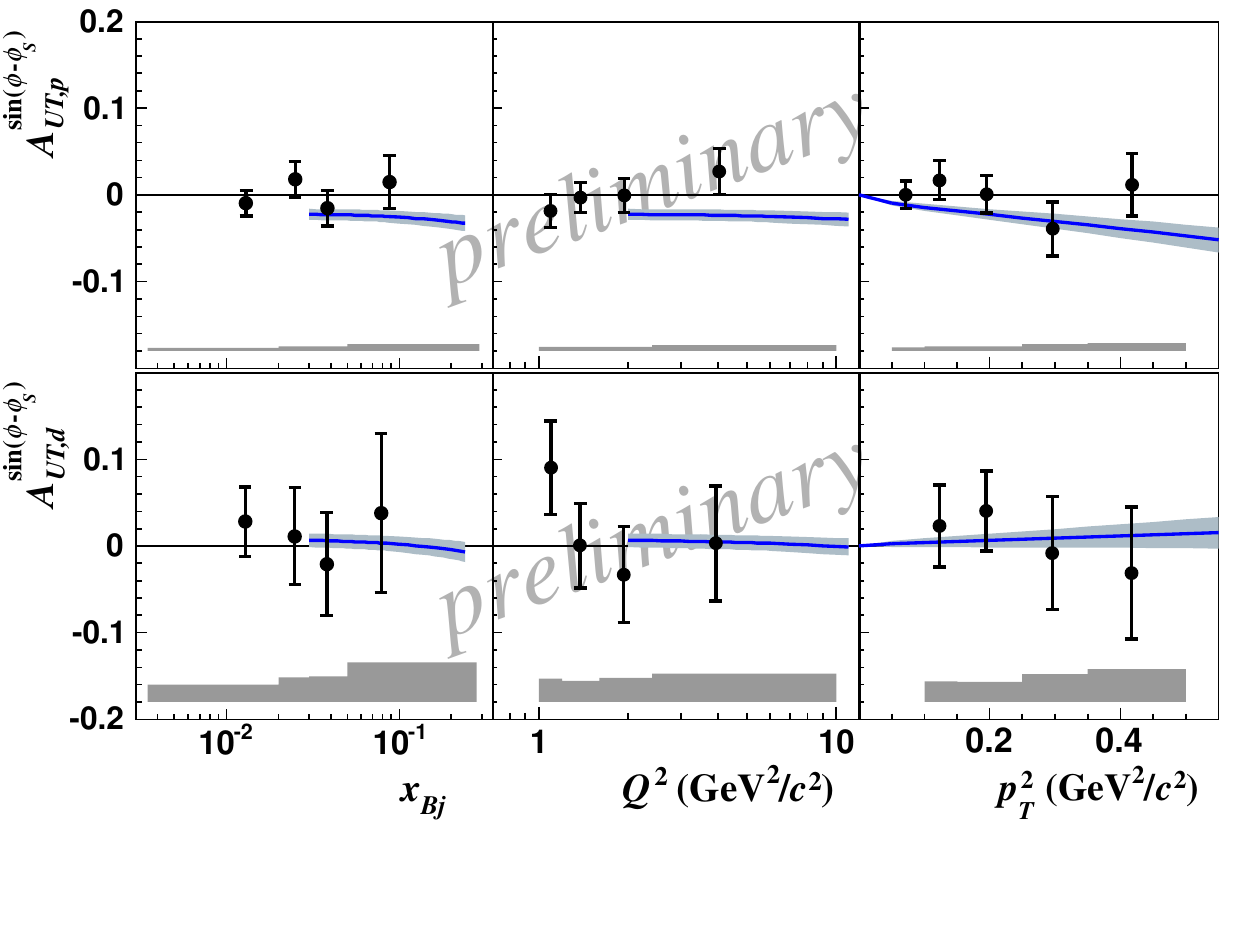}
\caption{Results for $A^{\sin(\phi-\phi_S)}_{UT}$ measured on proton (upper) and deuteron (lower) as a function of $x_{Bj}$, $Q^2$ and $p_T^2$. The systematic uncertainties are indicated by grey bands. The curves show the predictions of the GPD model~\cite{Goloskokov:2008ib} using the set of parameters called `variant 1'. The theoretical error bands take into account uncertainties of GPD parameterisations.}
\label{PIC:results}

\end{center}
\end{figure}

The results for the transverse target spin asymmetries
$A_{UT}^{\sin(\phi-\phi_S)}$ measured on proton and deuteron as a
function of $x_{Bj}$, $Q^2$ or $p_T^2$, are shown in
Fig.~\ref{PIC:results}. For both targets the asymmetries are found to be
small and compatible with zero within statistical uncertainties.
For transversely polarised deuterons it is the first measurement. The proton results are compatible with the results measured by HERMES~\cite{Rostomyan:2007rm}. Note that the COMPASS proton results are more precise by a factor of about 3 and cover a larger kinematic domain. The results are compared with predictions of the GPD model by Goloskokov and Kroll~\cite{Goloskokov:2008ib}, 
taking  into account only contributions from valence quark GPDs $E^u$ and $E^d$. A reasonable agreement is achieved. The small value of the asymmetry can be explained by an approximate cancellation of comparable contributions of opposite signs from GPDs $E^u$ and $E^d$: $E^u \approx -E^d$. For proton the asymmetry is sensitive to $2/3 E^u + 1/3 E^d$, while for deuteron it depends on $E^u + E^d$.

\bibliographystyle{DISproc}
\bibliography{wollny_heiner}

\end{document}